# The Influence of Age and Gender Information on the Diagnosis of Diabetic Retinopathy: Based on Neural Networks

Long Bai*, *Student Member, IEEE*, Sihang Chen*, Mingyang Gao*, Leila Abdelrahman, Manal Al Ghamdi, *Member, IEEE*, and Mohamed Abdel-Mottaleb, *Fellow, IEEE*

*Abstract*— This paper proposes the importance of age and gender information in the diagnosis of diabetic retinopathy. We utilized Deep Residual Neural Networks (ResNet) and Densely Connected Convolutional Networks (DenseNet), which are proven effective on image classification problems and the diagnosis of diabetic retinopathy using the retinal fundus images. We used the ensemble of several classical networks and decentralized the training so that the network was simple and avoided overfitting. To observe whether the age and gender information could help enhance the performance, we added the information before the dense layer and compared the results with the results that did not add age and gender information. We found that the test accuracy of the network with age and gender information was 2.67% higher than that of the network without age and gender information. Meanwhile, compared with gender information, age information had a better help for the results.

*Clinical Relevance*— The additional information in the dataset (such as age, gender, time of illness, etc.) can improve the accuracy of automatic diagnosis. Therefore, we strongly recommend that researchers add these different kinds of additional information when creating the dataset.

## I. INTRODUCTION

Diabetic retinopathy (DR) is a common cause of impaired vision in diabetic patients. With the further deepening of the world population's aging, diabetic retinopathy incidences have gradually increased. It has become one of the four major causes of blindness in Western countries [1, 2]. Analyzing the retinal fundus image can detect diabetic retinopathy as early as possible in the early stage of the disease, and intervene to alleviate the development of the disease, so the early identification and analysis of the retina images have become a focus of the current medical community.

### A. Neural Networks (NNs)

The neural network is a mathematical model that imitates animal neural networks' behavioral characteristics and performs distributed parallel information processing. It is the basis of deep learning. It is more flexible and has higher performance than traditional algorithms. With the development of computer technology, the development of neural network technology has gradually matured and is now widely used in pattern recognition, signal processing, and other fields. Neural network technology is common in the field of retinal image processing. J. David et al. [3] used the learning vector quantization network (LVQ) and backpropagation network to classify diabetic retinopathy by severity; Harry Pratt et al. [4] used the convolutional neural networks to distinguish whether the retina is suffering from DR; Shaohua Wan et al. [5] used VggNet, ResNet, GoogLeNet and other networks to identify DR; Saket S. Chaturvedi et al. [6] used DenseNet to complete the detection of diabetic retinopathy; Gargrya [7], Kevis-Kokitsi Maninis [8], Alvaro S. Hervella [9], and others have also done many attempts in this area.

### B. Multi-model Fusion

The existing neural network algorithms for directly detecting diabetic retinopathy input the retinal fundus image yet ignore the patient's other information corresponding to the image. However, H. A. Kahn [10], Ning Cheung [11], and others' research shows how diabetic retinopathy is related to the patient's gender, age, time of illness, and other information. Thus, we hope to improve the neural network's efficiency and accuracy in identifying diseases by adding the patient's gender and age information to the input of the neural network.

To introduce the influence of age and gender, we paid attention to the concept of multimodal fusion. Multimodal fusion refers to the technology in which the machine obtains information from multiple fields such as text, image, voice, and video. It then converts and merges the information in these different fields. Because the information in these different fields is often relevant to the situation that needs to be recognized, compared with classifying only one mode, multimodal fusion can improve the overall classification accuracy. It is currently widely used in speech recognition, sentiment analysis, video classification, and other fields. The main methods of multimodal fusion include joint representation, coordinated representation, encoder, and decoder, etc. [12-14] In this paper, we take the patients' age and gender information and the retinal image as the information in the two fields of text and image respectively. They represent the two fields that need to be fused. We used these two kinds of information as input for multimodal fusion to improve our model's performance.

### C. Multi-model Ensemble

Different models have different expression capabilities for data. The multi-model ensemble method can integrate the

*The first three authors contributed equally to the paper.

Long Bai is with the Beijing Institute and Technology, Beijing, China (corresponding author, phone: ; fax: ; e-mail: 1120171516@bit.edu.cn).

Sihang Chen is with the Beijing Institute and Technology, Beijing, China (e-mail: 1120170274@bit.edu.cn).

Mingyang Gao is with the Beihang University, Beijing, China (e-mail: sheepg@buaa.edu.cn).

Leila Abdelrahma is with the University of Miami, Coral Gables, FL 33124 USA (e-mail: leila.abdelrahman@miami.edu).

Manal Al Ghamdi, is with Umm Al-Qura University, Mecca. Saudi Arabia (e-mail: maalghamdi@uqu.edu.sa).

Mohamed Abdel-Mottaleb is with the University of Miami, Coral Gables, FL 33124 USA (e-mail: mottaleb@miami.edu).

advantages of different models to improve classification accuracy. Researchers input the same training set image into several different neural networks, such as ResNet [20] and DenseNet [21], and then merge the output to obtain the result. The final ensemble model is generally better than the sub-model [16-18]. Common ensemble methods include voting, averaging, bagging, boosting, stacking, etc.

## II. METHODOLOGY

### A. Adding the Age and Gender Information

Our goal was to build a neural network to identify whether a person has DR through a person's fundus images. Meanwhile, we hoped to add age and gender information to compare whether this information is helpful for the classification of diabetes. We first used several classic networks, specifically ResNet50, ResNet101, DenseNet121, DenseNet161, and DenseNet169, to train separately. Since we needed to use both eyes' data to make predictions together, we superimposed both eyes' images so that the number of input channels would be 6. We treated age and gender information, that is, text information, as multimodal and fuse them with retinal images, which are image information. Our preliminary experiments showed that the increase of the training accuracy was not obvious when we directly added age and gender information into the input, so we tried a simple but efficient fusion method. First, we used the fundus images as the input of the neural network alone and got an output, and then we put this output together with age and gender information into a new full connection layer for training. Figure 1 shows the method of adding age and gender information.

We used the above networks (ResNet and DenseNet) to train the Kaggle and OIA-ODIR datasets separately, without adding age and gender information. The accuracy on Kaggle's dataset [22] could reach nearly 90%, but the training accuracy on OIA-ODIR was less than 70%. Finally, we decided to use a neural network ensemble to increase training accuracy.

Meanwhile, we did not want to make the network too complicated and over-fitting. Thus, we did not choose traditional neural network ensemble methods but divided the entire training process into two parts. We also divided the training set into two parts (training set 1 and training set 2). We used training set 1 to train the above five traditional deep learning networks, respectively. Then we built five pre-trained models after each network training, so we could use each data in training set 2 as input in those pre-trained models, and got the output of five two-dimensional vectors. Then we merged the five vectors to form a ten-dimensional vector, took the several ten-dimensional vectors as input, used the label of the training set 2 as the actual value, and trained through the fully connected neural network. Finally, when testing the network's performance, we let the test set pass through the five pre-trained neural networks. We then merged all results into a matrix composed of ten-dimensional vectors, and the prediction results would be obtained through the trained fully connected layer. This method decentralized the training while avoiding become a more extensive neural network from the ensemble of traditional neural networks, and alleviated overfitting. Figure 2 shows the training and testing processes.

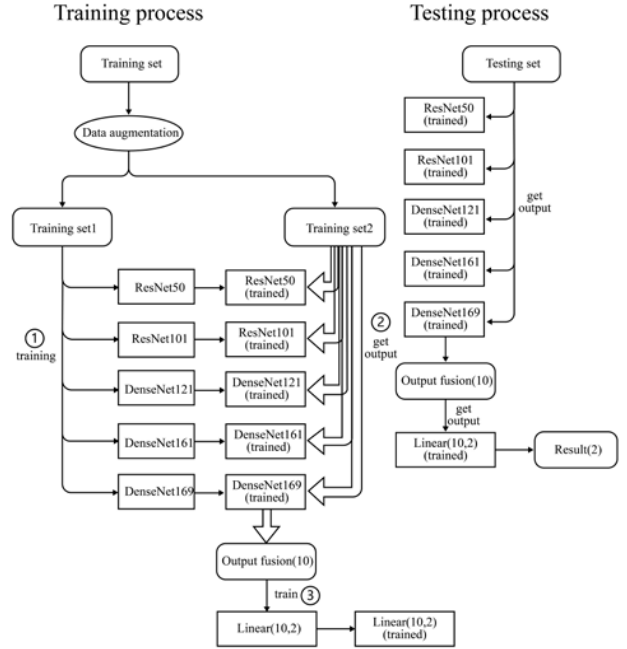

Figure 2. Training and testing process

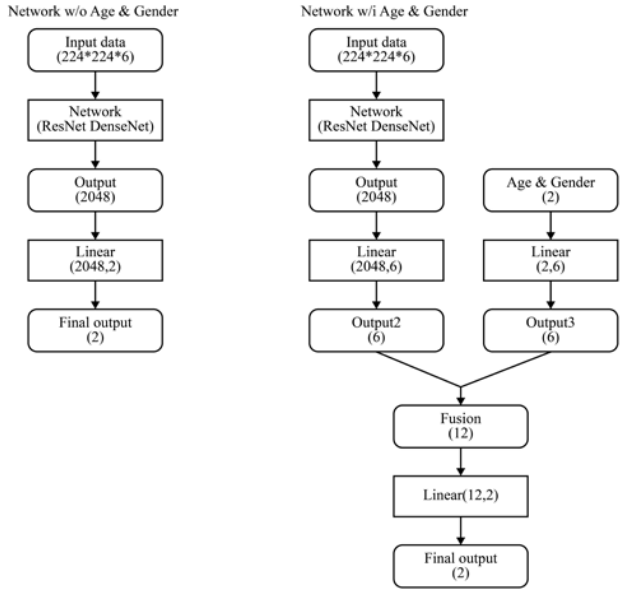

Figure 1. Comparison of the neural network with and without age and gender information

### B. Training and Testing

Unfortunately, we could hardly find a public dataset with age and gender information. Almost all the datasets do not include this information, so we had to choose the only dataset containing age and gender information, OIA-ODIR [15]. However, the image quality of this dataset is relatively low.

We choose cross-entropy as the loss function and run it on the GPU of Google Colaboratory Pro. The data preprocessing method is normalization, that is,

$$X = (X - Min) / (Max - Min) \qquad (1)$$

Meanwhile, we set the initial learning rate to 0.01, monitor the loss during training, and reduce the learning rate to 1/10 of the original when reducing loss is not apparent.

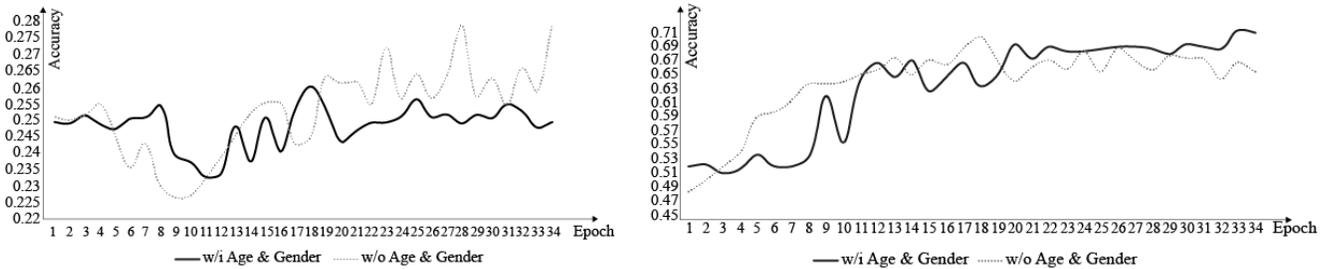

Figure 3: Tendency of loss and accuracy of ResNet50

TABLE I. ACCURACY OF TRAININGS USING DIFFERENT MODELS WITH DIFFERENT DATASTES

| Datasets | Neural Networks | | | | | |
|---|---|---|---|---|---|---|
| | ResNet50 | ResNet101 | DenseNet121 | DenseNet161 | DenseNet169 | Average |
| w/o Age & Gender | 0.6967 | 0.65 | 0.6833 | 0.6867 | 0.6733 | 0.678 |
| w/i Gender | 0.7076 | 0.6533 | 0.69 | 0.6967 | 0.68 | 0.6853 |
| w/i Age & Gender | 0.7133 | 0.6633 | 0.7033 | 0.7076 | 0.7167 | 0.7007 |
| w/i Age | 0.7233 | 0.6867 | 0.7067 | 0.7233 | 0.7167 | 0.7113 |

## III. EXPERIMENT VALIDATION

We conducted experiments on fundus images, divided into data with age and gender information and data without age and gender information, and conducted training and testing, respectively. We used the OIA-ODIR dataset and manually filtered the images based on the visible smudges in the images. In this dataset, each image had seven different disease labels, including diabetes. We only use the diabetes label. We divided the dataset into training set 1, training set 2, and test set. Meanwhile, we normalized the data matrix of age, gender, and image to the range of 0-1, respectively. As the CNN is a high-capacity model, we adopted the standard methods of rotation and flipping to enlarge the dataset to avoid problems such as overfitting caused by the dataset's limited size. When we added regularization to the network, the gradient learning rate method and early stop were also adopted to prevent overfitting.

### A. Training Respectively

We put training set 1 with age and gender information and training set 1 without age and gender information on the five networks ResNet50, ResNet101, DenseNet121, DenseNet161 and DenseNet169 respectively for training, and measured their accuracy. We used training set 1 as the training set and used the test set for testing. One of our models' training processes, specifically ResNet50, is shown in Figure 3 with the tendency of loss and accuracy of partial epochs.

The result shows in Table I that in the first training, the accuracy trained on the data with age and gender was 2.27% higher than the accuracy trained on the data without age and gender. Precision and recall also had better performance. This result could significantly reflect our work on integrating age and gender information and proved the influence of gender and age information on the prediction results.

Besides, we would like to know which age or gender has more influence on the accuracy of diagnosis. In the preliminary experiment, we input a number representing age information and a number representing gender information in the first training stage's fully connected layer. Next, we replaced the input with two identical numbers representing age information, and tested the results. Then we replaced them with gender and tested again. It turned out that gender had a slight improvement in accuracy, while age significantly improved the performance.

It is worth mentioning that in the first training session, the recall was relatively low. In the dataset, some patients will have diabetes symptoms in one eye and healthy features in the other eye. However, during training, they were put into the network for training together. The labels were all marked as negative samples, that is, samples suffering from diabetes, which would weaken the model's ability to learn positive examples, leading to such a situation.

### B. Model Ensemble Training

In the second stage of training, we passed the training set 2 through the model trained by training set 1, and put the obtained output into a fully connected layer for training. Next, we put the results on the test set for testing. Table II shows the first training results.

TABLE II. ACCURACY OF TRAININGS USING MULTI-MODEL ENSEMBLE WITH DIFFERENT DATASETS

| Datasets | Training Stages | | |
|---|---|---|---|
| | Training1 | Training2 | Diff. |
| w/o Age & Gender | 0.678 | 0.72 | 0.042 |
| w/i Gender | 0.6853 | 0.7233 | 0.038 |
| w/i Age & Gender | 0.7007 | 0.7467 | 0.046 |
| w/i Age | 0.7113 | 0.7533 | 0.042 |

We compared the two training results, and concluded that, in the four sets of experiments, the accuracy increased by 4.2%, 3.8%, 4.6%, 4.2% respectively, after the five models' ensemble. Figure 4 shows these results.

Simultaneously, after the model ensemble, models trained

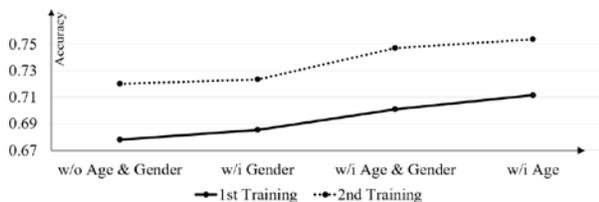
Figure 4: Tendency of accuracy of the final results

with age and gender information also had higher performance than models trained without age and gender information: multimodal model accuracy is higher by 2.67%. We think this is enough to prove the importance of age and gender information in diabetic retinopathy diagnosis. Meanwhile, according to the comparative experiment in the second column of Table II, we can see that the effect of age information on accuracy is much greater than that of gender information.

It is worth mentioning that we extracted a small part of the data (300 images) from Kaggle Diabetic Retinopathy Detection 2015 for two-category classification training of a simple network - ResNet50. The result was a relatively high accuracy, which could reach more than 88%. However, when we extracted the same number of images from the OIA-ODIR dataset for training, the accuracy was only about 67%. If we could find a diabetic retinopathy dataset of fundus images with both high-quality images, and additional information (age, gender, etc.), our accuracy will not be limited to this but will be significantly improved. We strongly recommend that researchers bring age, gender, and other relevant information when making relevant datasets, which can significantly improve our auto-diagnosis accuracy.

## IV. CONCLUSION AND DISCUSSION

We proposed the importance of age and gender information in the diagnosis of diabetic retinopathy. We used an ensemble of several classic networks, introduced ResNet and DenseNet implementations, which have made outstanding achievements in the field of computer vision, into the diagnosis of diabetic retinopathy, and added age and gender information as multimodal for training. We verified it on the OIA-ODIR dataset, and the experimental results show that our method of network ensemble and our method of fusing multimodal information has dramatically improved the accuracy.

The diabetic retinopathy diagnosis aided by artificial intelligence can significantly help ophthalmologists make a preliminary diagnosis of the disease. In the future, we plan to use methods such as transfer learning [23] to train on high-quality datasets in advance to improve the overall accuracy.

ACKNOWLEDGMENT

Thanks to Google Colab Pro for providing GPU for deep learning training. We are very grateful to Yanheng Li, School of Design and Arts, Beijing Institute of Technology for helping us draw the graphs and charts in the paper.